\documentclass[twocolumn,showpacs,amsmath,amssymb,prb]{revtex4}

\usepackage{graphicx}   	
\usepackage{dcolumn}    	
\usepackage{bm}         	
\newcommand{\rd}{{\rm d}}
\newcommand{\ri}{{\rm i}}
\newcommand{\re}{{\rm e}}
\newcommand{\kc}{k_{\rm c}}
\newcommand{\Fp}{F_{\rm p}}

\begin{document}

\title[Generalized acceleration theorem]
      {Generalized acceleration theorem for spatiotemporal Bloch waves}

\author{Stephan Arlinghaus}

\author{Martin Holthaus}

\affiliation{Institut f\"ur Physik, Carl von Ossietzky Universit\"at,
	D-26111 Oldenburg, Germany}

\date{July 29, 2011}

\begin{abstract}
A representation is put forward for wave functions of quantum particles in 
periodic lattice potentials subjected to homogeneous time-periodic forcing, 
based on an expansion with respect to Bloch-like states which embody both 
the spatial and the temporal periodicity. It is shown that there exists
a generalization of Bloch's famous acceleration theorem which grows out of
this representation, and captures the effect of a weak probe force applied 
in addition to a strong dressing force. Taken together, these elements point 
at a ``dressing and probing'' strategy for coherent wave-packet manipulation, 
which could be implemented in present experiments with optical lattices. 
\end{abstract}




\maketitle


\section{Introduction}

The so-called acceleration theorem for wave-packet motion in periodic
potentials, formulated already in 1928 by Bloch,~\cite{Bloch28} 
has proven to be of outstanding value to solid-state physics for 
understanding the dynamics of Bloch electrons within a semiclassical 
picture.~\cite{AshcroftMermin76,Kittel87} In its most-often used variant, 
this theorem states that if we consider an electronic wave packet in a 
spatially periodic lattice, which is centered in $k$ space around some wave 
vector $\vec{k}_{\rm c}$, and if an external electric field $\vec{E}(t)$ is 
applied under single-band conditions, then this center wave vector evolves 
in time according to $\hbar\dot{\vec k}_{\rm c}(t) = -e\vec{E}(t)$, with 
$-e$ being the electronic charge. Perhaps its best-known application is the 
explanation of Bloch oscillations of particles exposed to a homogeneous, 
constant force,~\cite{RoskosEtAl92,FeldmannEtAl92,Feldmann92,WaschkeEtAl93,
IgnatovEtAl94,RotvigEtAl95,IgnatovEtAl95,BouchardLuban95,UnterrainerEtAl96, 
LyssenkoEtAl97,LenzEtAl99,BenDahanEtAl96,MorschEtAl01}
which we recapitulate here in the simplest guise: Take a particle in a 
one-dimensional tight-binding energy band $E(k) = -(W/2)\cos(ka)$, where 
$W$ is the band width and $a$ denotes the lattice period. Assume that the 
particle's wave packet is centered around $\kc (0)$ initially and subjected 
to a homogeneous force of strength~$F$. Then the acceleration theorem, now
taking the form
\begin{equation} 
	\hbar\dot{k}_{\rm c}(t) = F \; ,
\label{eq:OAT}	
\end{equation}
tells us $\kc (t) = \kc (0) + Ft/\hbar$, so that the packet moves through 
$k$ space at a constant rate.~\cite{Bloch28} According to another classic 
work by Jones and Zener,~\cite{JonesZener34} the particle's group velocity 
$v_{\rm g}(t)$ in real space is determined, quite generally, by the derivative 
of $E(k)$ with respect to $k$ when evaluated at the moving center $\kc (t)$,     
\begin{equation}
	v_{\rm g}(t) 
	= \frac{1}{\hbar} \left. \frac{\rd E}{\rd k} \right|_{\kc (t)} \; .
\label{eq:OGV}
\end{equation}	
In our case, this relation immediately gives
\begin{equation}	
	v_{\rm g}(t) 	
	= \frac{Wa}{2\hbar} \sin\!\big(\kc (0)a + \omega_{\rm B} t\big) \; ,
\end{equation}
implying that the particle's response to the constant force is an oscillating
motion with the Bloch frequency~\cite{Zener34} $\omega_{\rm B} = Fa/\hbar$. 
This elementary example, to which we will come back later in Sec.~\ref{sec:S_4},
strikingly illustrates the power of this type of approach. But an obvious 
restriction stems from the necessity to remain within the scope of the 
single-band approximation; the above acceleration theorem~(\ref{eq:OAT}) is 
put out of action when several Bloch bands are substantially coupled by the 
external force. Nonetheless, in the present work we demonstrate that there 
exists a generalization of the acceleration theorem which can be applied even 
under conditions of strong interband transitions. Specifically, we consider 
situations in which a Bloch particle is subjected to a strong oscillating 
force which possibly induces pronounced transitions between the unperturbed 
energy bands. By abandoning the customary crystal-momentum 
representation~\cite{Callaway74} and introducing an alternative Floquet 
representation instead, we show that the effect of an additional force then is 
well captured by another acceleration theorem which closely mimics the spirit
of the original. We obtain two major results: The Floquet analog~(\ref{eq:GAT})
of Bloch's acceleration theorem~(\ref{eq:OAT}), and the Floquet 
analog~(\ref{eq:FGV}) of the Jones-Zener expression~(\ref{eq:OGV}) for 
the group velocity. These findings are particularly useful for control 
applications, when a strong oscillating field ``dresses'' the lattice and thus 
significantly alters its band structure, while a second, comparatively weak 
homogeneous force is employed to effectuate controlled population transfer. 
We first outline the formal mathematical arguments in  Secs.~\ref{sec:S_2}
and \ref{sec:S_3}, and then we give two applications of topical interest, 
discussing ``super'' Bloch oscillations in Sec.~\ref{sec:S_4} and coherently 
controlled interband population transfer in Sec.~\ref{sec:S_5}. Although we 
restrict ourselves here for notational simplicity to one-dimensional lattices, 
our results can be carried over to general, higher-dimensional settings.

\section{The Floquet representation}
\label{sec:S_2}    

We consider a particle of mass~$m$ moving in a one-dimensional lattice
potential $V(x) = V(x+a)$ with spatial period~$a$ under the influence
of a homogeneous, time-dependent force~$F(t)$, as described by the
Hamiltonian
\begin{equation}
	\widetilde{H}_0(x,t) = \frac{p^2}{2m} + V(x) - F(t)x \; .
\label{eq:HOR}
\end{equation}
Subjecting the particle's wave function $\widetilde{\psi}(x,t)$ to the 
unitary transformation
\begin{equation}
	\widetilde{\psi}(x,t) = \exp\!\left(
	\frac{\ri}{\hbar} x \! \int_0^t \! \rd \tau \, F(\tau) \right) 
	\psi(x,t) \; ,	
\label{eq:UNT}
\end{equation}	  
the new function $\psi(x,t)$ obeys the Schr\"odinger equation
\begin{equation}
	\ri\hbar \frac{\partial}{\partial t} \psi(x,t) =
	H_0(x,t) \psi(x,t) \; ,
\label{eq:TSE}
\end{equation}
with the transformed Hamiltonian
\begin{equation}
	H_0(x,t) = 
	\frac{1}{2m}\left(p + \int_0^t \! \rd \tau \, F(\tau) \right)^2
	+ V(x) \; .
\label{eq:UPH}	
\end{equation}
Now let us further assume that the force $F(t)$ is periodic in time with 
period~$T$, such that its one-cycle integral either vanishes or
equals an integer multiple of $\hbar$ times the reciprocal lattice wave
number $2\pi/a$:
\begin{equation}
	\int_0^T \! \rd t \, F(t) = r \times \hbar \frac{2\pi}{a}
	\quad , \quad  r = 0,\pm 1,\pm 2,\ldots \; .
\label{eq:RES}
\end{equation}
For example, this is accomplished by a monochromatic oscillating force 
with an additional static bias,
\begin{equation}
	F(t) = F_r + F_{\rm ac}\cos(\omega t) \; ,
\label{eq:INT}
\end{equation}
provided the latter satisfies the condition $F_r a = r\hbar\omega$. 
Then the Floquet theorem guarantees that the time-dependent Schr\"odinger 
equation~(\ref{eq:TSE}) admits a complete set of spatiotemporal Bloch 
waves,~\cite{Holthaus92,DreseHolthaus97,ArlinghausHolthaus10} 
that is, of solutions of the form
\begin{equation}
	\psi_{n,k}(x,t) = 
	\exp\!\big( \ri k x - \ri \varepsilon_n(k) t/\hbar \big)
	u_{n,k}(x,t) \; ,
\label{eq:STB}
\end{equation}
with spatially {\em and\/} temporally periodic functions 
\begin{equation}
	u_{n,k}(x,t) = u_{n,k}(x+a,t) = u_{n,k}(x,t+T) \; . 
\label{eq:DPU}
\end{equation}   
As usual, $n$ is the band index and $k$ a wave number; $\varepsilon_n(k)$
thus is the quasienergy dispersion relation for the $n$th band. If $r = 0$ 
in Eq.~(\ref{eq:RES}), the existence of these solutions is obvious, 
because then $H_0(x,t) = H_0(x+a,t) = H_0(x,t+T)$, so that the wave 
functions~(\ref{eq:STB}) generalize the customary Bloch waves~\cite{Bloch28} 
for particles in spatially periodic lattice potentials by also accounting for 
the temporal periodicity of the driving force. When $r \neq 0$, so that 
$H_0(x,t)$ itself is not periodic in time, spatiotemporal Bloch 
waves~(\ref{eq:STB}) emerge nonetheless because $k$ is projected to the first 
quasimomentum Brillouin zone, as first discussed by Zak.~\cite{Zak93} In any 
case, the quasienergies $\varepsilon_n(k)$ may depend in a complicated manner 
on the parameters of the driving force, and the wave functions 
$\psi_{n,k}(x,t)$ pertaining to a single quasienergy band may be nontrivial 
mixtures of several unperturbed energy bands. For later use, we observe that 
their spatial parts
\begin{equation}
	\varphi_{n,k}(x,t) = \exp(\ri k x) u_{n,k}(x,t)
\label{eq:DFP}
\end{equation}     
obey the quasienergy eigenvalue equation
\begin{equation}
	\left( H_0(x,t) - \ri\hbar\frac{\partial}{\partial t} \right)
	\varphi_{n,k}(x,t) = \varepsilon_n(k) \varphi_{n,k}(x,t) \; ,
\label{eq:QEE}
\end{equation}
as follows immediately when plugging the solutions~(\ref{eq:STB}) into the 
Schr\"odinger equation~(\ref{eq:TSE}). Throughout, we adopt the standard
normalization
\begin{equation}
	\int_{-\infty}^{\infty} \! \rd x \, 
	\varphi^*_{n',k'}(x,t) \varphi_{n,k}(x,t)
	= \frac{2\pi}{a} \delta_{n,n'} \delta(k-k') \; .
\label{eq:NRM}
\end{equation} 
An arbitrary wave packet $\psi(x,t)$ may now be expanded with respect to 
these spatiotemporal Bloch waves, and written in the form
\begin{equation}
	\psi(x,t) = \sum_n \sqrt{\frac{a}{2\pi}} \int_{\cal B} \! \rd k \, 
	g_n(k,t) \varphi_{n,k}(x,t) \; ,
\label{eq:FLR}	
\end{equation}
with ${\cal B} = [-\pi/a, \pi/a[$ denoting the fundamental Brillouin
zone. The expansion coefficients $g_n(k,t)$ depend on the way the system
has been prepared and on the way the driving force has been turned on,
whereas the basis functions $\varphi_{n,k}(x,t)$ and their quasienergies
$\varepsilon_n(k)$ are given by the eigenvalue equation~(\ref{eq:QEE}) and 
obviously are independent of such details. Clearly, one has
\begin{equation}
	g_n(k,t) = g_n(k,0) \exp\!\big(-\ri\varepsilon_n(k)t/\hbar \big) \; ,
\label{eq:DGT}	
\end{equation}
so that the populations $|g_n(k,t)|^2$ remain constant in time. This 
expansion~(\ref{eq:FLR}), referred to as the Floquet representation of the
wave packet, is formally reminiscent of its customary crystal-momentum 
representation, that is, of an expansion with respect to the Bloch states 
of the unperturbed potential $V(x)$ which underlies the standard acceleration 
theorem.~\cite{Callaway74,GrecchiSacchetti01} There are, however, substantial 
differences which become most clear when considering a wave packet occupying 
a single quasienergy band,
\begin{equation}
	\psi(x,t) = \sqrt{\frac{a}{2\pi}} \int_{\cal B} \! \rd k \, 
	g(k,t) \varphi_{k}(x,t) \; ;	
\label{eq:SBP}	
\end{equation}   		 
here and in the following, we omit the band index~$n$ for ease of notation.
Now this wave packet~(\ref{eq:SBP}) may describe, for instance, the dynamics 
in a situation where two unperturbed energy bands are resonantly coupled by 
the driving force $F(t)$; consequently, in a crystal-momentum representation 
one would have to account for Rabi-type oscillations between these two bands
by coefficients which quantify the oscillating band populations. In the 
Floquet respresentation, on the other hand, the Rabi oscillations are already 
incorporated into the basis states~(\ref{eq:STB}), so that one merely 
encounters single quasienergy band dynamics, with the remaining time evolution 
of $g(k,t)$ simply given by Eq.~(\ref{eq:DGT}). Thus, although the external 
force effectuates transitions between the unperturbed Bloch bands, there are 
no inter-quasienergy band transitions; $|g(k,t)|^2$ remains constant in time. 
Second, even in a situation where $F(t)$ does  not couple different energy 
bands, the wave packet's center $\kc (t)$ evolves according to the standard 
acceleration theorem $\hbar\dot{k}_{\rm c} = F$ in the crystal-momentum 
representation, whereas in the Floquet representation the moment
\begin{equation}
	\langle k \rangle = \int_{\cal B} \! \rd k \, k | g(k,t) |^2
\label{eq:FRA}
\end{equation}  
obviously stays constant in time. In short, an expansion of the wave packet 
with respect to the spatiotemporal Bloch waves~(\ref{eq:STB}) implies 
constant coefficients, and hence constant occupation probabilities, if the 
external force $F(t)$ adheres to the specification~(\ref{eq:RES}). This formal 
shift of the dynamics from the occupation numbers to the basis states which is 
implied by the Floquet representation now allows for a clear and physically 
transparent description of the additional effects which emerge when the
external force does {\em not\/} obey Eq.~(\ref{eq:RES}); these effects
are captured by the generalized acceleration theorem exposed in the following.

\section{The Floquet acceleration theorem}
\label{sec:S_3}

We take a wave packet occupying a single quasienergy band and stipulate that 
in addition to the possibly strong driving force~$F(t)$ there is a second 
homogeneous force $\Fp (t)$ which we denote as the {\em probe force\/}; this 
is assumed to be sufficiently weak so that it does not introduce transitions 
among different quasienergy bands. To be precise, the total Hamiltonian now 
reads
\begin{equation}
	\widetilde{H}(x,t) = \frac{p^2}{2m} + V(x) - F(t)x - \Fp (t)x \; , 
\label{eq:HOT}
\end{equation} 
where the time-periodic force $F(t)$ is resonant in the sense of 
Eq.~(\ref{eq:RES}) and thus creates a basis of spatiotemporal Bloch
waves~(\ref{eq:STB}), whereas the probe force $\Fp (t)$ also is spatially 
homogeneous, but not necessarily periodic in time. After performing the 
unitary transformation~(\ref{eq:UNT}), we obtain the Hamiltonian in the form
\begin{equation}
	H(x,t) = H_0(x,t) - \Fp (t) x \; ,
\end{equation}
with $H_0(x,t)$ given by Eq.~(\ref{eq:UPH}). Moreover, we start from an 
initial wave packet of the form~(\ref{eq:SBP}). Because of the additional 
probe force $\Fp (t)$, the time evolution of $g(k,t)$ is no longer given by 
Eq.~(\ref{eq:DGT}); the aim now is to find an effective Hamiltonian ${\cal H}$ 
which governs the resulting dynamics of $g(k,t)$, under the proposition that 
this remains restricted to the single, initially occupied quasienergy band.       

Exploiting the normalization~(\ref{eq:NRM}), we have
\begin{equation}
	g(k,t) = \sqrt{\frac{a}{2\pi}} \int \! \rd x \,   
	\varphi_k^*(x,t) \psi(x,t) \; .
\end{equation}
This gives 
\begin{eqnarray}
	\ri\hbar\frac{\partial g}{\partial t} & = &   
	\sqrt{\frac{a}{2\pi}} \int \! \rd x 
	\left( \ri\hbar \frac{\partial \varphi_k^*}{\partial t} \psi
	+ \varphi_k^* H \psi \right)
\nonumber\\	& = &
	\sqrt{\frac{a}{2\pi}} \int \! \rd x 
	\left( 
	\left[ H_0 - \ri\hbar\frac{\partial}{\partial t}\right] 
	\varphi_k \right)^* \psi 		
\nonumber\\	& &
	- \sqrt{\frac{a}{2\pi}} \Fp \int \! \rd x \,
	\varphi_k^* x \psi \; , 	
\end{eqnarray}	 
having suppressed the arguments $x$ and $t$ for better legibility; all
integrals here are taken over the entire lattice. In the first term on 
the right-hand side of this equation we exploit the quasienergy eigenvalue 
equation, Eq.~(\ref{eq:QEE}), yielding $\varepsilon(k)g(k,t)$. For rewriting 
the second term we use
\begin{equation}
	\varphi_k^* x = \ri \partial_k \varphi_k^*
	- \ri \re^{-\ri k x} \partial_k u_k^* \; ,
\end{equation}
which is obtained by taking the derivative of the complex conjugate to 
Eq.~(\ref{eq:DFP}) with respect to $k$, and leads to 
\begin{eqnarray}	   
	\sqrt{\frac{a}{2\pi}} \int \! \rd x \, \varphi_k^* x \psi 
	& = & 	
	\ri \partial_k g - \ri \sqrt{\frac{a}{2\pi}} \int \! \rd x \, 
	\re^{-\ri k x} \partial_k u_k^* \psi
\nonumber \\	& = &
	\ri \partial_k g - \ri \langle \partial_k u_k | u_k \rangle g \; .  	
\end{eqnarray}	
For making the final step, we have resubstituted the 
expression~(\ref{eq:SBP}) for $\psi$ and have made use of the identity
\begin{equation}
	\int \! \rd x \, \re^{\ri(k'-k)x} u_{k'} \partial_k u_k^*
	= \frac{2\pi}{a}\delta(k-k') \langle \partial_k u_k | u_k \rangle \; ,
\end{equation}
with the scalar product
\begin{equation}
	\langle \partial_k u_k | u_k \rangle = 
	\int_0^a \! \rd x \, u_k(x,t) \partial_k u_k^* (x,t)
\label{eq:DEF}
\end{equation}
being given by an integral over a single lattice period. Note that
\begin{equation}
	\langle u_k | u_k \rangle = 1 \; ,
\label{eq:NRU}
\end{equation}
as an immediate consequence of Eq.~(\ref{eq:NRM}), which implies
\begin{equation}
	\langle\partial_k u_k | u_k \rangle +
	\langle\partial_k u_k | u_k \rangle^* = 0 \; ,
\end{equation}
so that $\langle\partial_k u_k | u_k \rangle$ is purely imaginary.
Collecting all the pieces, we obtain the desired evolution equation
\begin{equation}
	\ri \hbar \frac{\partial}{\partial t} g(k,t)
	= {\cal H} g(k,t) \; ,
\label{eq:EEG}
\end{equation}
with the effective Hamiltonian for the Floquet representation,
\begin{equation}
	{\cal H} = \varepsilon(k) - \ri \Fp \partial_k 
	- \Fp \, {\rm Im} \langle\partial_k u_k | u_k \rangle \; . 
\label{eq:EFH}
\end{equation} 
From this expression we deduce the generalized acceleration theorem, 
that is, the acceleration theorem for the Floquet representation: Since 
the moment~(\ref{eq:FRA}) obeys the the equation
\begin{equation}
	\frac{\rd}{\rd t}\langle k \rangle = 
	\frac{\ri}{\hbar}\langle [{\cal H}, k] \rangle  	 
\end{equation}
and the commutator appearing here on the right-hand side is easily evaluated,
$\ri[{\cal H}, k] = \Fp$, we are directly led to
\begin{equation}
	\hbar \frac{\rd}{\rd t}\langle k \rangle(t) = \Fp (t) \; .	
\label{eq:GAT}
\end{equation}
This is the central result of the present work; its analogy to the standard
acceleration theorem~(\ref{eq:OAT}) for the crystal-momentum representation is 
evident. Observe that there is an intuitively clear reason for the appearance 
of the term proportional to $\langle\partial_k u_k | u_k \rangle$ in the 
effective Hamiltonian~(\ref{eq:EFH}): The twofold periodic parts $u_k(x,t)$ 
of the spatiotemporal Bloch waves are obtained by solving the eigenvalue
equation~(\ref{eq:QEE}). This is done for each wave number~$k$ separately,
so that one is free to bestow upon each eigensolution an arbitrary phase
factor $\exp(\ri\theta(k))$. On the other hand, the evolution 
equation~(\ref{eq:EEG}) for the wave function $g(k,t)$ in the Floquet 
representation naturally establishes a ``connection'' between those
different eigensolutions~\cite{Simon83,Berry84} and therefore requires
information about the gauge function $\theta(k)$; this is provided by
the expression $\langle\partial_k u_k | u_k \rangle$. Note further that when 
multiplying Eq.~(\ref{eq:EEG}) by $g^*(k,t)$ and subtracting the complex 
conjugate of the resulting equation, this piece drops out, and one is left
with
\begin{equation}
	\left( \frac{\partial}{\partial t} 
	+ \frac{\Fp (t)}{\hbar} \frac{\partial}{\partial k} \right)
	| g(k,t) |^2 = 0 \; .
\end{equation}  
Thus, $| g(k,t) |^2$ does not depend on $k$ and $t$ separately, but rather 
on the combination $k - \int_0^t \! \rd \tau \, \Fp (\tau)/\hbar$, so that 
the distribution $g(k,t)$ moves through the Floquet $k$ space without change of 
shape, again in precise analogy to the classic behavior.~\cite{Bloch28} But 
we reemphasize that this seemingly simple dynamics might be unrecognizable 
in the usual crystal-momentum representation, because the system might undergo 
violent transitions between different energy bands when monitored in a basis 
of time-independent Bloch waves.

As the introductory example has shown, the standard (crystal-momentum) 
acceleration theorem develops its main power in combination with the 
Jones-Zener expression~(\ref{eq:OGV}) for the wave packet's group velocity 
in real space, and the question naturally arises whether there exists a 
similar connection in the Floquet representation. Obviously, one can establish 
a relation corresponding to Eq.~(\ref{eq:OGV}) by applying a stationary-phase 
argument to the expansion~(\ref{eq:FLR}), but here we follow an alternative 
line of reasoning which may be found particularly enlightening. Considering a 
well-localized wave packet $\widetilde{\psi}(x,t)$ in the original frame of 
reference to which the Hamiltonian operators~(\ref{eq:HOR}) and (\ref{eq:HOT}) 
pertain, that packet's group velocity is given by      
\begin{eqnarray}
	v_{\rm g}(t) & = &\frac{\rd}{\rd t} 
	\langle \widetilde{\psi}(x,t) | x | \widetilde{\psi}(x,t) \rangle
\nonumber\\	& = &
	\frac{1}{m}
	\langle \widetilde{\psi}(x,t) | p | \widetilde{\psi}(x,t) \rangle 
	\; .		
\label{eq:VGD}
\end{eqnarray}
On the other hand, exploiting the operator identity
\begin{equation}
	\re^{-ikx} p \re^{ikx} = p + \hbar k \; , 
\label{eq:UDO}
\end{equation}
the eigenvalue equation~(\ref{eq:QEE}) transforms into the even more basic 
eigenvalue equation
\begin{equation}
	\left( H_k(x,t) - \ri\hbar\frac{\partial}{\partial t} \right)
	u_{n,k}(x,t) = \varepsilon_n(k) u_{n,k}(x,t) 
\label{eq:EPU}
\end{equation}
for the periodic core pieces $u_{n,k}(x,t)$ of the spatiotemporal Bloch 
waves~(\ref{eq:STB}), invoking the parametrically $k$-dependent operator
\begin{equation}
	H_k(x,t) = \frac{1}{2m}
	\left(p + \hbar k + \int_0^t \! \rd \tau \, F(\tau) \right)^2
	+ V(x) \; . 	
\end{equation}
This eigenvalue problem can efficiently be implemented for numerical 
calculations.~\cite{DreseHolthaus97} It also manifestly contains the origin
of the condition~(\ref{eq:RES}) imposed on the oscillating force~$F(t)$, 
since $k$ is reduced to fall within ${\cal B}$. Most importantly, this 
eigenvalue problem~(\ref{eq:EPU}) poses itself in an extended Hilbert space
made up of functions $u_{n,k}(x,t)$ which are periodic in both space and 
time, in accordance with Eq.~(\ref{eq:DPU}). Consequently, ``time'' has
to be regarded as a coordinate in this extended Hilbert space and therefore
needs to be integrated over when forming a scalar product, just like any
spatial coordinate. Thus, the natural scalar product in this extended Hilbert
space is given by~\cite{Sambe73}    
\begin{equation}
	\langle\!\langle \, \cdot \, | \, \cdot \, \rangle\!\rangle
	\equiv \frac{1}{T} \int_0^T \! \rd t \, 
	\langle \, \cdot \, | \, \cdot \, \rangle \; ,
\label{eq:SCP}
\end{equation}
with $\langle \, \cdot \, | \, \cdot \, \rangle$ denoting the standard
scalar product in the original, physical Hilbert space, as already employed
in Eqs.~(\ref{eq:DEF}) and (\ref{eq:NRU}). It follows that the quasienergies 
$\varepsilon_n(k)$ can be written as diagonal elements of the matrix of the 
quasienergy operator, 
\begin{equation}
	\varepsilon_n(k) = 
	\langle\!\langle u_{n,k} | 
	H_k - \ri\hbar\partial_t| u_{n,k} \rangle\!\rangle \; ,
\end{equation}
inviting us to make use of an analog of the Hellmann-Feynman 
theorem:~\cite{Sambe73}
\begin{eqnarray}
	\frac{\rd}{\rd k} \varepsilon_n(k) & = & \langle\!\langle u_{n,k} | 
	\frac{\rd H_k}{\rd k} | u_{n,k} \rangle\!\rangle
\nonumber \\	& = &
	\frac{\hbar}{m}\langle\!\langle u_{n,k} |
	p + \hbar k + \int_0^t \! \rd \tau \, F(\tau) |
	u_{n,k} \rangle\!\rangle
\nonumber \\	& = &
	\frac{\hbar}{m}\langle\!\langle \widetilde{\psi}_{n,k} | p |
	\widetilde{\psi}_{n,k} \rangle\!\rangle \; .
\label{eq:VGE}		
\end{eqnarray}		 	
In the final step we have undone the shift~(\ref{eq:UDO}); the wave functions 
$\widetilde{\psi}_{n,k}(x,t)$ then denote the functions which are obtained
from the spatiotemporal Bloch waves~(\ref{eq:STB}) by inverting the
transformation~(\ref{eq:UNT}). Comparison of Eqs.~(\ref{eq:VGD}) and 
(\ref{eq:VGE}), keeping in mind the definition~(\ref{eq:SCP}), now yields 
the desired relation: Supposing that $\widetilde{\psi}(x,t) =
\widetilde{\psi}_{n,k_0}(x,t)$ were made up from a single spatiotemporal 
Bloch wave labeled by $n$ and $k_0$, say, one would obtain the formal 
identity
\begin{equation}
	\overline{v}_{\rm g} 
	\equiv \frac{1}{T} \int_0^T \! \rd t \, v_{\rm g}(t)
	= \frac{1}{\hbar} \left. 
	\frac{\rd \varepsilon_n}{\rd k} \right|_{k_0} \; .
\end{equation}
But this is not what we want, because an individual spatiotemporal Bloch 
wave is uniformly extended over the lattice and thus does not correspond 
to a ``group'' which propagates in space. Rather, we require a wave 
packet~(\ref{eq:SBP}) which is reasonably well centered in the
Floquet $k$ space, with a center $\langle k \rangle$ given by 
Eq.~(\ref{eq:FRA}). Then we have
\begin{equation}
	\overline{v}_{\rm g} = \frac{1}{\hbar} \left. 
	\frac{\rd \varepsilon_n}{\rd k} \right|_{\langle k \rangle} 
\label{eq:FGV}
\end{equation}
to good accuracy, so that the cycle-averaged group velocity of the Floquet 
wave packet is given by the derivative of its quasienergy dispersion 
relation, evaluated at its center $\langle k \rangle$. Again, this Floquet 
relation~(\ref{eq:FGV}) closely mimics its historic crystal-momentum 
antecessor, given by Eq.~(\ref{eq:OGV}). In contrast to the equation of 
motion~(\ref{eq:GAT}) for $\langle k \rangle$ itself, which holds exactly 
within a single quasienergy band setting, this relation~(\ref{eq:FGV}) is 
an approximation which holds the better, the narrower the packet's Floquet 
$k$ space distribution. Although it seems self-evident, it might be worthwhile 
to stress that the argument required to evaluate the derivative~(\ref{eq:FGV}) 
is Floquet $\langle k \rangle$, not crystal momentum $\kc$.

\section{Super Bloch oscillations}
\label{sec:S_4}

The phenomenon termed ``super'' Bloch oscillations~\cite{HallerEtAl10} arises
when a Bloch particle is subjected to both a static (dc) and an oscillating
(ac) force, such that an integer multiple of the ac frequency is only slightly 
detuned from the Bloch frequency associated with the dc component of the 
force.~\cite{HallerEtAl10,KolovskyKorsch10,AlbertiEtAl09,KudoMonteiro11}   
Although the effect itself appears almost trivial from the mathematical point 
of view, we nonetheless dwell on this at some length, because it provides a 
particularly instructive example for juxtaposing the familiar crystal-momentum 
representation to the Floquet representation introduced in Sec.~\ref{sec:S_2}
and for demonstrating in detail how they match. To be definite, we consider
the total force to be of the form 
\begin{equation}
	F(t) = \Theta(t-t_0) 
	\left[ F_{\rm dc} + F_{\rm ac}\cos(\omega t) \right] \; ,
\label{eq:TTO}
\end{equation}
where $\Theta(t)$ denotes the Heaviside function, so that both the dc and
the ac component of the force are turned on instantaneously and simultaneously
at $t_0$; that moment $t_0$ thus determines the relative phase between the
Bloch oscillations caused by the dc component and the driving oscillations of 
the ac component.  

The basic assumptions now are that {\em (i)\/} we are given an initial wave 
packet which occupies a single energy band, being centered around $\kc (t_0)$ 
at the moment $t = t_0$ in the crystal-momentum representation, and that 
{\em (ii)\/} interband transitions remain negligible for $t > t_0$, despite 
the action of the force $F(t)$. We then encounter single-band dynamics 
which are fully covered by the ``old'' acceleration theorem 
$\hbar \dot{k}_{\rm c}(t) = F(t)$, 
giving    
\begin{eqnarray}
	\kc (t) & = & \kc (t_0) 
\\	       	& + & \frac{1}{\hbar}\left[ F_{\rm dc}(t - t_0)
		+ \frac{F_{\rm ac}}{\omega}\sin(\omega t)
		- \frac{F_{\rm ac}}{\omega}\sin(\omega t_0) \right] 
\nonumber		
\end{eqnarray}
for $t > t_0$. As an archetypal example we now take a tight-binding cosine 
energy dispersion relation for the band considered,  
\begin{equation}
	E(k) = -\frac{W}{2}\cos(ka) \; ,
\end{equation}
parametrized as in the Introduction. Utilizing Eq.~(\ref{eq:OGV}),
one then finds the packet's group velocity:
\begin{equation}
	v_{\rm g}(t) 
	= \frac{1}{\hbar} \left. \frac{\rd E}{\rd k} \right|_{\kc (t)}
	= \frac{Wa}{2\hbar} \sin\!\big( \kc (t)a \big) \; . 
\label{eq:IGV}
\end{equation}
This expression describes super Bloch oscillations if we assume further
that the dc component of the force is almost resonant in the sense of
Eq.~(\ref{eq:RES}). We therefore decompose this component according to 
\begin{equation}
	F_{\rm dc} = F_r + \delta F \; ,
\label{eq:DEC}
\end{equation}
where $F_r a = r \hbar\omega$ with some nonzero integer~$r$ as previously 
in Eq.~(\ref{eq:INT}), so that $r\omega$ equals the Bloch frequency 
$\omega_{\rm B} = F_ra/\hbar$, while $\delta F$ is quite small compared to 
$F_r$. We then have
\begin{equation}
	F_{\rm dc}a = r \hbar\omega + \hbar\delta\omega \; ,	
\end{equation}
with frequency detuning $\delta\omega = \delta F a/\hbar$, so that the
group velocity~(\ref{eq:IGV}) takes the form 
\begin{equation}
	v_{\rm g}(t) 
	= \frac{Wa}{2\hbar}\sin\!\Big( r\omega t + \delta\omega \, t 
	+ K\sin(\omega t) + \Phi \Big) \; ,
\end{equation}
having introduced the scaled driving amplitude
\begin{equation}
	K = \frac{F_{\rm ac}a}{\hbar\omega}
\label{eq:SDA}
\end{equation}
and a constant phase 
\begin{equation}
	\Phi = \kc (t_0)a - (r\omega + \delta\omega)t_0 - K\sin(\omega t_0) \; ,
\label{eq:PHI}
\end{equation}	
which accounts for the initial conditions. Because $\delta\omega \ll \omega$
according to our specifications, the contribution $\delta\omega\, t$ to the
argument of $v_{\rm g}$ does not vary appreciably during one single cycle
$T = 2\pi/\omega$ of the ac component. Thus, when averaging the instantaneous
group velocity over one such cycle, this ``slow'' time dependence may be 
ignored, meaning that $\delta\omega\, t$ may be considered as constant when
taking the average.~\cite{KudoMonteiro11} Invoking the Jacobi-Anger indentity
in the guise
\begin{equation}
	\re^{\ri K \sin(\omega t)} = \sum_{\ell=-\infty}^\infty
	{\rm J}_\ell(K) \re^{\ri\ell\omega t} \; ,
\end{equation}	
where ${\rm J}_\ell(K)$ denote the Bessel functions of the first kind,
one immediately obtains 
\begin{eqnarray}
	\overline{v}_{\rm g}(t) & = & 
	\frac{1}{T} \int_0^T \! \rd t \, v_{\rm g}(t)
\nonumber \\	& = &	
	(-1)^r {\rm J}_r(K) \frac{Wa}{2\hbar}
	\sin(\delta\omega \, t + \Phi) \; .
\label{eq:CAC}
\end{eqnarray}
According to the above reasoning, here the ``fast'' time dependence 
is integrated out, but the slow dependence on $\delta\omega\, t$ 
remains.~\cite{KudoMonteiro11} Integrating, this yields the cycle-averaged 
drift motion of the packet, that is, its position $\overline{x}_{\rm g}(t)$ 
without the fast ac quiver, 
\begin{equation}
	\overline{x}_{\rm g}(t) = -(-1)^r {\rm J}_r(K) \frac{W}{2\delta F}
	\cos(\delta\omega \, t + \Phi) \; , 
\label{eq:SBO}
\end{equation}
with a suitably chosen origin of the $x$ axis. This result finally clarifies 
what is ``super'' with these dynamics: Because the residual force $\delta F$ 
is quite small, the amplitude of this oscillation~(\ref{eq:SBO}) can be 
fairly large; indeed, in a corresponding experiment with weakly interacting 
Bose-Einstein condensates in driven optical lattices Haller {\em et al.\/}
have observed giant center-of-mass oscillations with displacements across 
hundreds of lattice sites.~\cite{HallerEtAl10} 

As far as the phenomenon itself is concerned there is nothing more to add;
because one requires single Bloch-band dynamics right from the outset, a
Floquet treatment is not necessary. Nevertheless the Floquet approach is
of its own intrinsic value even here, since it provides a somewhat different 
view which, in contrast to the above crystal-momentum calculation, is capable 
of some generalization. 
   
The Floquet analysis starts from the spatiotemporal Bloch waves and their
quasienergies. In a single-band setting with an external homogeneous force,
these are exceptionally easy to obtain: Writing the Bloch waves of the 
undriven lattice in the form
\begin{equation}
	\varphi_k(x) = \sum_\ell w_\ell(x) \re^{\ri k \ell a} \; ,  
\label{eq:BLO}
\end{equation}
where $w_\ell(x)$ denotes a Wannier function localized around the $\ell$th 
lattice site,~\cite{Kohn59} the so-called Houston functions~\cite{Houston40}
\begin{equation}
	\widetilde{\psi}_k(x,t) = 
	\sum_\ell w_\ell(x) \re^{\ri q_k(t)\ell a}
	\exp\!\left( -\frac{\ri}{\hbar}\int_0^t \! \rd \tau \, 
	E\!\big( q_k(\tau) \big) \right)  
\label{eq:HOU}	
\end{equation}
are solutions to the time-dependent Schr\"odinger equation in the original 
frame, for arbitrary $F(t)$, provided the ``moving wave numbers'' $q_k(t)$
are given by  
\begin{equation}
	q_k(t) = k + \frac{1}{\hbar}\int_0^t \! \rd \tau \, F(\tau) \; ,
\label{eq:QKT}
\end{equation}
always assuming the viability of the single-band 
approximation.~\cite{EckardtEtAl09} Taking a force of the particular 
form~(\ref{eq:INT}) with exactly resonant $F_r$ obeying $F_r a = r\hbar\omega$,  
we have	
\begin{equation}	
	q_k(t) = k + \frac{r\omega t}{a} 
	+ \frac{F_{\rm ac}}{\hbar\omega}\sin(\omega t) \; .
\end{equation}	
This implies that both the exponentials $\exp\!\big(\ri q_k(t)\ell a\big)$
and $E\big(q_k(t)\big)$ are $T$ periodic in time, with $T = 2\pi/\omega$,
whereas the integral over $E\big(q_k(t)\big)$ is not, because the Fourier
expansion of $E\big(q_k(t)\big)$ contains a zero mode, so that its integral
contains a linearly growing contribution. But this observation reveals that 
the ``accelerated Bloch waves''~(\ref{eq:HOU}) with resonant time-periodic
forcing~(\ref{eq:INT}) are precisely the required spatiotemporal Bloch waves 
in the original frame, with their quasienergies being determined by the 
zero mode: 
\begin{eqnarray}
	\varepsilon(k) & = & \frac{1}{T}\int_0^T \! \rd t \,
	E\!\big( q_k(t) \big)
\nonumber \\	& = &
	-(-1)^r {\rm J}_r(K) \frac{W}{2} \cos(k a) \; .
\label{eq:QED}
\end{eqnarray}
The remarkable fact that the quasienergy bands collapse, {\em i.e.\/}, become 
flat when $K$ is such that ${\rm J}_r(K) = 0$, indicates that an oscillating 
force can effectively shut down the tunneling contact between neighboring wells; 
this ``coherent destruction of tunneling'' is a generic feature of driven
single-band systems.~\cite{EckardtEtAl09,GrifoniHanggi98,Longhi05,KudoEtAl09} 
A bit of reflection then shows that the core pieces $u_k(x,t)$ of the
spatiotemporal Bloch waves, that is, the solutions to the eigenvalue
equation~(\ref{eq:EPU}), are given by
\begin{eqnarray} 
	u_k(x) & = & \sum_{\ell} w_\ell(x) \re^{\ri k \ell a}
\nonumber \\	& \times &		
	\exp\!\left( -\frac{\ri}{\hbar}\int_0^t \! \rd \tau \, 
	\left[ E\!\big( q_k(\tau) \big) - \varepsilon(k) \right] \right) \; .  
\end{eqnarray}
Although this has not been particularly emphasized, the above construction 
makes sure that any spatiotemporal Bloch wave~(\ref{eq:HOU}) is labeled by 
the same wave number $k$ as the ordinary Bloch wave to which it reduces when 
the external force vanishes.~\cite{EckardtEtAl09} Otherwise, there is nothing 
particular about the choice $t = 0$ for the lower bound of integration 
in Eq.~(\ref{eq:QKT}) for $q_k(t)$: In contrast to Eq.~(\ref{eq:TTO}),
where $t = t_0$ has been singled out as the moment when the force is turned 
on, and which thus designates an initial-value problem for a particular wave 
packet, the solution of the eigenvalue problem~(\ref{eq:EPU}) for the entire
spatiotemporal Bloch basis requires a force $F(t)$ which is perfectly 
periodic in time; the resulting expression for $q_k(t)$ thus holds for both 
$t > 0$ and $t < 0$. Also note that it would be meaningless to include some 
additional constant phase into the argument of the ac component of the 
force~(\ref{eq:INT}): Because this expression holds for all times~$t$, such 
a phase would merely amount to a shift of the origin of the time coordinate 
and thus is as irrelevant for the calculation of the quasienergy dispersion 
relation as would be a shift of the origin of the spatial coordinate system 
for the calculation of a crystal's energy band structure.

Knowing the quasienergy dispersion relation~(\ref{eq:QED}), the machinery
established in Sec.~\ref{sec:S_3} can be set to work: According to 
Eq.~(\ref{eq:FGV}), the cycle-averaged group velocity of a Floquet wave 
packet~(\ref{eq:SBP}) is given by        
\begin{eqnarray}
	\overline{v}_{\rm g} & = & \frac{1}{\hbar}
	\left. \frac{\rd \varepsilon}{\rd k} \right|_{\langle k \rangle}
\nonumber \\	& = &
	(-1)^r {\rm J}_r(K) \frac{Wa}{2\hbar} \sin(\langle k \rangle a) \; .	
\label{eq:CAG}
\end{eqnarray}	
If we now turn back to the specific forcing~(\ref{eq:TTO}), and thus consider
exactly the same initial-value problem as in the previous crystal-momentum
exercise, we can make operational use of the decomposition~(\ref{eq:DEC}) of 
the dc force: Its resonant part $F_r$ has already been incorporated into the 
spatiotemporal Bloch waves~(\ref{eq:HOU}), which means that it has already 
been accounted for in ``dressing'' the lattice and changing its original 
energy dispersion $E(k)$ to the quasienergy dispersion $\varepsilon(k)$. 
Therefore, it is only the small residual part $\delta F$ which enters into 
the equation of motion for $\langle k \rangle$, that is, into the generalized 
acceleration theorem~(\ref{eq:GAT}); this part $\delta F$ thus constitutes
a particular, time-independent example of a probe force $\Fp (t)$ as considered 
in Sec.~\ref{sec:S_3}. We now have
\begin{equation}
	\hbar \frac{\rd}{\rd t}\langle k \rangle(t) 
	= \frac{\hbar\delta\omega}{a} \; ,
\label{eq:PAR}
\end{equation}	
giving
\begin{equation}
	\langle k \rangle(t) = \langle k \rangle(t_0)
	+ \frac{\delta\omega}{a}(t - t_0) \; .	
\label{eq:FKT}
\end{equation}
All that remains to be done now is to express the initial Floquet center
$\langle k \rangle(t_0)$ in terms of the initial wave packet's center 
$\kc (t_0)$, which had been specified in the crystal-momentum representation. 
But this is an easy task, comparing the original Bloch waves~(\ref{eq:BLO}) to 
their spatiotemporal descendents~(\ref{eq:HOU}): At the moment $t_0$ when the 
force~(\ref{eq:TTO}) is turned on, $\kc (t_0)$ coincides with $q_k(t_0)$ for 
one particular $k$; this evidently is the desired $\langle k \rangle(t_0)$.
The equality identifying $\langle k \rangle(t_0)$ thus is
\begin{equation}
	\kc (t_0) = q_{\langle k \rangle(t_0)}(t_0) \; ,
\end{equation}
which, written out in full detail, reads
\begin{equation}
	\kc (t_0) = \langle k \rangle(t_0) + \frac{r\omega t_0}{a}
	+ \frac{F_{\rm ac}}{\hbar\omega}\sin(\omega t_0) \; .
\end{equation}	
Using this to eliminate $\langle k \rangle(t_0)$ from Eq.~(\ref{eq:FKT}),
we arrive at
\begin{eqnarray}
	\langle k \rangle(t) a & = & \kc (t_0)a - r\omega t_0 
	- K\sin(\omega t_0) + \delta\omega(t - t_0)
\nonumber \\	& = &
	\delta \omega \, t + \Phi		 
\label{eq:ARG}
\end{eqnarray}	
with precisely the same phase $\Phi$ as already defined in 
Eq.~(\ref{eq:PHI}). Inserting this argument~(\ref{eq:ARG}) into the 
cycle-averaged group velocity~(\ref{eq:CAG}), and comparing with the previous 
expression~(\ref{eq:CAC}), one confirms that the result of the Floquet analysis 
fully coincides with that of the more customary crystal-momentum calculation. 
The necessity to painstakingly distinguish between crystal momentum~$\kc$ and 
Floquet~$\langle k \rangle$ at all stages may appear a bit mind-boggling; if 
this is not done with sufficient care, one might overlook a contribution to 
$\Phi$.~\cite{KudoMonteiro11} But if respected properly, the mathematical 
structure of the Floquet picture unerringly leads to the correct answer.     

If one strips the above reasoning to the bare essentials, that is, if one 
starts from the quasienergy dispersion relation~(\ref{eq:QED}), takes its
derivative to obtain the formal expression~(\ref{eq:CAG}) for the 
cycle-averaged group velocity, and then inserts the solution to the equation
of motion~(\ref{eq:PAR}) dictated by the generalized acceleration theorem in 
order to compute the group velocity of the wave packet actually considered, 
one sees that this procedure exactly parallels the explanation of the usual 
Bloch oscillations, as reviewed in the Introduction. Thus, super Bloch
oscillations may be seen as ordinary Bloch oscillations arising in response 
to a weak probe force $\delta F$, but occurring in a spatiotemporal lattice, 
as created by dressing the original lattice through application of the strong 
force~(\ref{eq:INT}).

One might finally wish to get away from the particular, instantaneous onset
of the forcing assumed in Eq.~(\ref{eq:TTO}): The dc and the ac component
might not be switched on simultaneously, or not abruptly, possibly involving
two different turn-on functions for the two components. In any case, at
some moment $t_0$ the final amplitudes will have been reached, so that the
previous analysis goes through unaltered for $t > t_0$, if one only interprets 
$\kc (t_0)$ correctly: This would no longer indicate the crystal-momentum wave 
number around which the initial wave packet had been prepared, but rather that 
to which the latter had been shifted during the turn-on phase. Expressed 
differently, the phase $\Phi$ in Eqs.~(\ref{eq:CAC}) and (\ref{eq:SBO}) depends 
significantly on the precise turn-on protocol: Not surprisingly, the way the 
external force has been turned on in the past crucially affects the 
coherent wave-packet motion after the turn-on is over.                

Aside from its aesthetic value, the Floquet picture offers at least one 
further benefit: Bloch oscillations in dressed lattices may also occur under 
conditions such that the quasienergy bands are mixtures of several unperturbed 
energy bands, disabling a crystal-momentum treatment. A Floquet analysis, 
on the other hand, would merely require one to replace the single-band 
quasienergies~(\ref{eq:QED}) by the actual ones and then again invoke 
the generalized acceleration theorem~(\ref{eq:GAT}), similar to the examples 
worked out in the next section.

\section{Coherent control of interband population transfer}
\label{sec:S_5}

A field of major current interest in which the Floquet picture may find 
possibly groundbreaking applications concerns ultracold atoms, or weakly
interacting Bose-Einstein condensates, in time-periodically driven optical
lattices.~\cite{DreseHolthaus97,HallerEtAl10,AlbertiEtAl09,EckardtEtAl09,
ZenesiniEtAl09,EckardtEtAl10,StruckEtAl11}
As opposed to ordinary crystalline matter exposed to high-power laser fields,
such systems offer the advantage that one can apply even nonperturbatively 
strong driving forces without inducing unwanted inhomogeneities, as caused 
by polarization effects or domain formation.~\cite{ArlinghausHolthaus10} The 
issue at stake here is not merely redoing well-known condensed-matter physics 
in another setting, and thus selling old wine in new skins, but rather finding 
genuinely new ways of coherently controlling mesoscopic matter waves, such 
that target states are created which have not been accessible before, and are 
manipulated according to some prescribed protocol. Here we point out that 
the generalized acceleration theorem~(\ref{eq:GAT}) may be a valuable tool 
in this quest.  
 
A standard one-dimensional (1D) optical lattice is described by a cosine
potential
\begin{equation}
	V(x) = \frac{V_0}{2}\cos(2k_{\rm L}x) \; ,
\label{eq:OLA}
\end{equation}
where $k_{\rm L}$ is the wave number of the two counterpropagating laser
beams generating the lattice.~\cite{MorschOberthaler06,BlochEtAl08} Its 
depth~$V_0$ is measured in multiples of the single-photon recoil energy 
\begin{equation}
	E_{\rm r} = \frac{\hbar^2 k_{\rm L}^2}{2m} \; .
\end{equation}
For orientation, if one traps $^{87}$Rb atoms in a lattice with $k_{\rm L}$
corresponding to the wavelength $\lambda = 842$~nm, as in a recent 
experiment by Zenesini {\em et al.\/},~\cite{ZenesiniEtAl09} one finds
$E_{\rm r} = 1.34 \times 10^{-11}$~eV; typical optical lattices are a few 
recoil energies deep.

\begin{figure}[t]
\centering
\includegraphics[width = 0.7\linewidth]{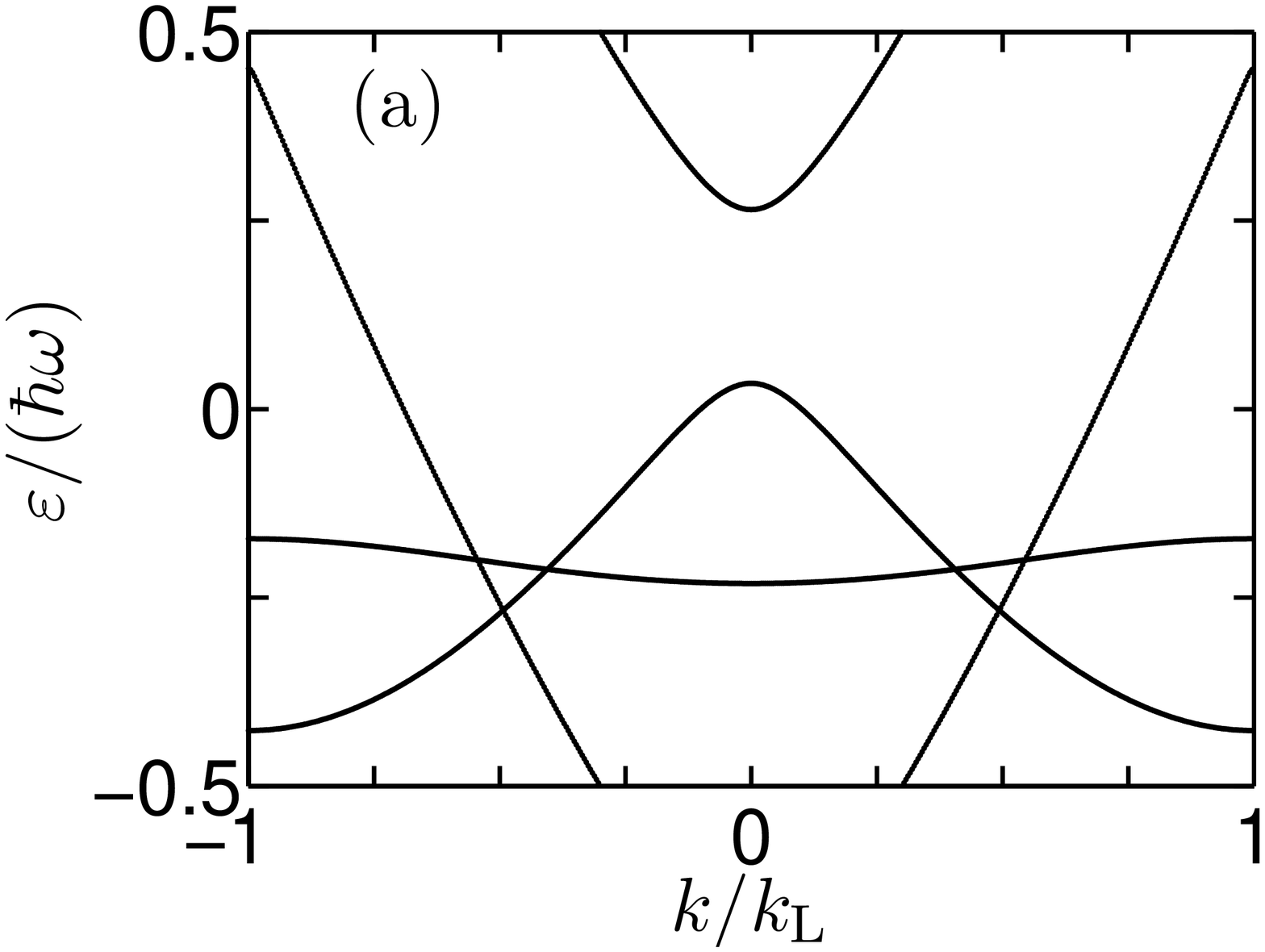}
\includegraphics[width = 0.7\linewidth]{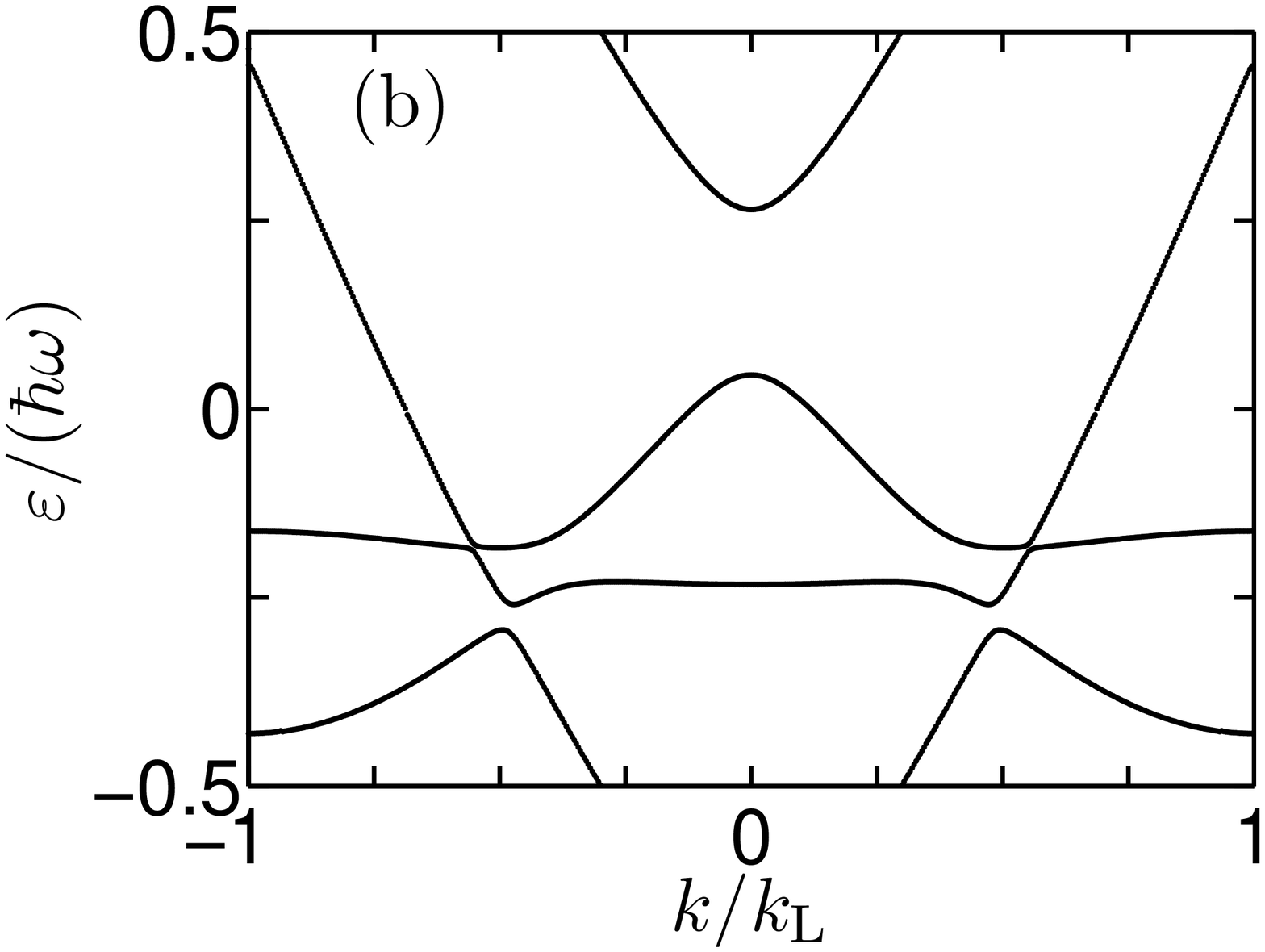}
\includegraphics[width = 0.7\linewidth]{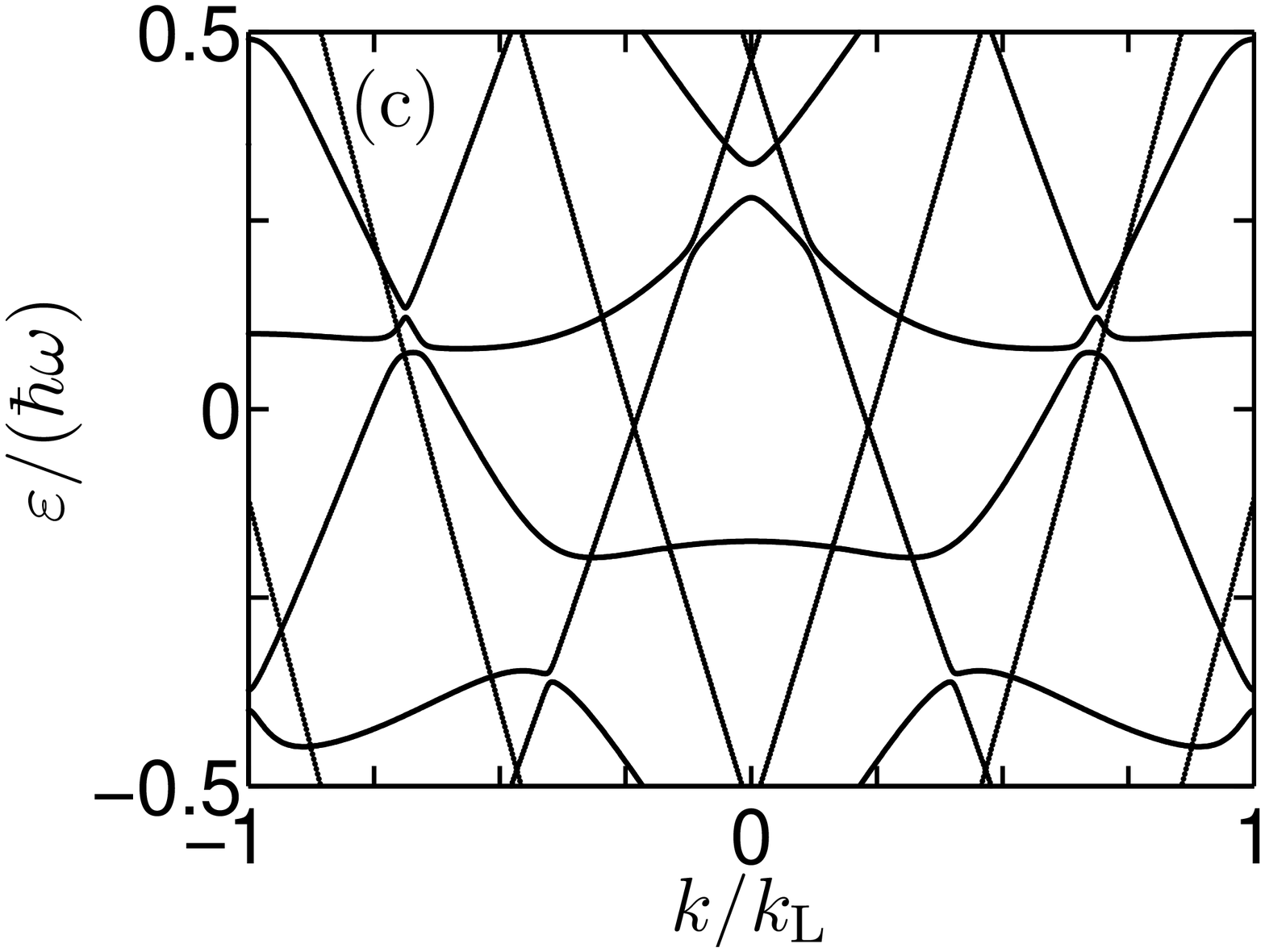}
\caption{(a) Quasienergy spectrum of an ac-driven 1D optical 
	lattice~(\ref{eq:OLA}) with depth $V_0/E_{\rm r} = 5.7$, scaled 
	driving frequency $\hbar\omega/E_{\rm r} = 3.71$, and scaled 
	driving amplitude $K = 0$. This figure is obtained by projecting 
	the lowest three energy bands of the undriven lattice to the 
	first quasienergy Brillouin zone, ranging from 
	$\varepsilon/(\hbar\omega) = -1/2$ to 
	$\varepsilon/(\hbar\omega) = +1/2$.
	(b) Quasienergy band structure for $K = 0.5$. Here the ac Stark
	shifts still are comparatively weak, but the time-periodic forcing
	introduces pronounced avoided crossings among the ``lowest'' three
	bands.
	(c) Quasienergy band structure for $K = 3.0$ for the ``lowest''
	five bands, revealing substantial ac Stark shifts.} 
\label{fig:F_1}
\end{figure}

Figure~\ref{fig:F_1} shows quasienergy spectra for such a 1D cosine 
lattice~(\ref{eq:OLA}) with depth $V_0/E_{\rm r} = 5.7$ under pure ac forcing, 
that is, for $F(t) = F_{\rm ac}\cos(\omega t)$ not containing a dc component,
with driving frequency $\omega = 3.71 \, E_{\rm r}/\hbar$. Under the 
laboratory  conditions specified above ($^{87}$Rb at $\lambda = 842$~nm), 
this corresponds to $\omega/(2\pi) = 12$~kHz. Figure~\ref{fig:F_1}(a) results 
when the scaled driving amplitude~(\ref{eq:SDA}) is set to zero; this subfigure
therefore is obtained by projecting the lowest three unperturbed energy 
bands to the fundamental quasienergy Brillouin zone, which extends from
$\varepsilon = -\hbar\omega/2$ to $\varepsilon = +\hbar\omega/2$ on the 
ordinate. Figure~\ref{fig:F_1}(b) displays the quasienergy band structure for 
the moderate driving strength $K = 0.5$; here avoided crossings show up which 
generally indicate multiphotonlike resonances.~\cite{ArlinghausHolthaus10} 
Figure~\ref{fig:F_1}(c) then reveals pronounced ac Stark shifts (that is, 
shifts of the quasienergies against the zone-projected original energies) 
for $K = 3.0$, corresponding to truly strong forcing.

\begin{figure}[t]
\centering
\includegraphics[width = 0.8\linewidth]{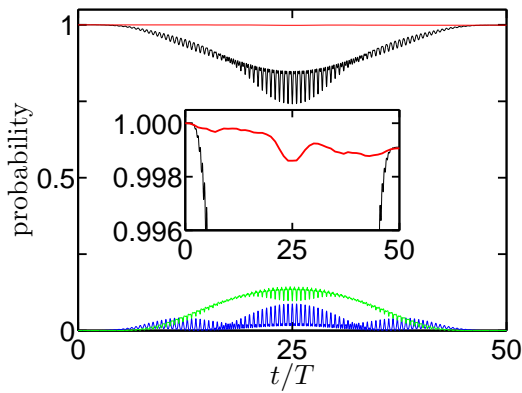}
\caption{(Color online) Time evolution of an initial wave packet~(\ref{eq:INI})
	under the action of a driving pulse~(\ref{eq:PUL}) with the smooth 
	squared-sine envelope~(\ref{eq:ENV}), with maximum scaled 
	amplitude $K_{\rm max} = 3.0$, scaled driving frequency 
	$\hbar\omega/E_{\rm r} = 3.71$, and pulse length  
	$T_{\rm p} = 50 \, T$, where $T = 2\pi/\omega$ is the duration of a
	single cycle. The main frame shows the occupation probabilities of 
	the original Bloch energy bands during the pulse. Apart from the 
	initially populated lowest band $n = 1$ (jagged line at the top), both 
	bands $n = 2$ and $n = 3$ become significantly excited during the 
	pulse (jagged lines at the bottom), the band $n = 3$ even to a higher 
	extent than $n = 2$ at maximum driving strength. In contrast, when 
	monitoring the same dynamics within the bases provided by the 
	instantaneous spatiotemporal Bloch waves, {\em only\/} the 
	corresponding Floquet band $n = 1$ is appreciably occupied, as 
	shown by the horizontal line at the top and magnified in the 
	inset. Observe the scale of the inset's ordinate!}    
\label{fig:F_2}
\end{figure}

We now turn from the quasienergy spectrum to an exemplary initial-value 
problem: At $t = 0$ we prepare an initial wave packet~(\ref{eq:SBP}) in the 
lowest Bloch band $n = 1$ with a Gaussian momentum distribution,
\begin{equation}
	g_1(k,0) = (\sqrt{\pi} \Delta k)^{-1/2}
	\exp\!\left(-\frac{k^2}{2(\Delta k)^2} \right) \; ,
\label{eq:INI}
\end{equation}
centered around $\kc (0)/k_{\rm L} = 0$ with width $\Delta k/k_{\rm L} = 0.1$,
and subject it to a pulse,
\begin{equation}
	F(t) = F_{\rm max} s(t) \sin(\omega t) \; ,	
\label{eq:PUL}	
\end{equation}
starting at $t = 0$ and ending at $t = T_{\rm p}$, endowed with a smooth,
squared-sine envelope function:
\begin{equation}
	s(t) = \sin^2\!\left(\frac{\pi t}{T_{\rm p}}\right)
	\quad , \quad  0 \le t \le T_{\rm p} \; .
\label{eq:ENV}	
\end{equation}
We again set $\omega = 3.71 \, E_{\rm r}/\hbar$, as in Fig.~\ref{fig:F_1};
adjust the pulse length to 50 cycles; $T_{\rm p} = 50 \times 2\pi/\omega$,
and fix the maximum driving amplitude $F_{\rm max}$ such that
$K_{\rm max} = F_{\rm max}a/(\hbar\omega) = 3.0$, corresponding to the
conditions reached in Fig.~\ref{fig:F_1}(c). We then monitor the resulting
wave-packet dynamics both in the basis of the unperturbed energy bands and in 
the bases provided by the instantaneous spatiotemporal Bloch waves, that is, 
in the family of Floquet bases which are obtained when the driving amplitude 
is kept fixed at any value $F_{\rm ac} = F_{\rm max}s(t_0)$ reached during 
the pulse. Figure~\ref{fig:F_2} displays the results: The jagged lines in the 
main frame show the occupation probabilities of the lowest three unperturbed 
Bloch bands $n = 1,2$, and 3; in the middle of the pulse the band $n = 3$ contains 
even more population than the band $n = 2$. On the other hand, the horizontal line at 
the top depicts the occupation of the instantaneous Floquet band emerging 
from the lowest Bloch band: This Floquet band contains practically {\em all\/} 
the population during the entire pulse, which means that the wave function 
adjusts itself adiabatically to the changing morphology of its quasienergy 
band,~\cite{ArlinghausHolthaus10} as previously sketched in 
Fig.~\ref{fig:F_1}, when the driving amplitude $F_{\rm max} s(t)$ is first
increased and then decreased back to zero. To quantify the precise degree 
of adiabatic following, the inset in Fig.~\ref{fig:F_2} shows the variation
of the Floquet band population on a much finer scale. Observe that the
final adiabaticity defect is on the order of merely $0.1\%$, even though 
the driving amplitude reaches its fairly high maximum strength within no more
than 25 cycles.

With respect to the concepts developed in Sec.~\ref{sec:S_2}, 
Fig.~\ref{fig:F_2} strikingly demonstrates the advantages of the Floquet 
picture over the traditional crystal-momentum representation for the situation
considered. If there were an additional probe force, its effect would have 
to be tediously disentangled from the fast oscillations of the Bloch band 
populations. When the same dynamics are seen from the Floquet viewpoint, 
essentially ``nothing'' happens, because practically all inter-Bloch-band 
transitions are already accounted for by continuously adapting the Floquet 
basis, so that the action of a probe force would stand out most clearly. 
Although, of course, the crystal-momentum representation is mathematically 
equivalent to the Floquet picture, there is no question which one is 
preferable here. Note also that Fig.~\ref{fig:F_2} answers one further 
pertinent question: How do we prepare a wave packet which occupies merely a 
single quasienergy band, although it is undergoing rapid transitions between 
several Bloch bands at the same time? The recipe for achieving this is simple: 
Start with a wave packet occupying a single Bloch band and switch on the 
driving force smoothly thereby enabling adiabatic following.

\begin{figure}[t]
\centering
\includegraphics[width = 0.8\linewidth]{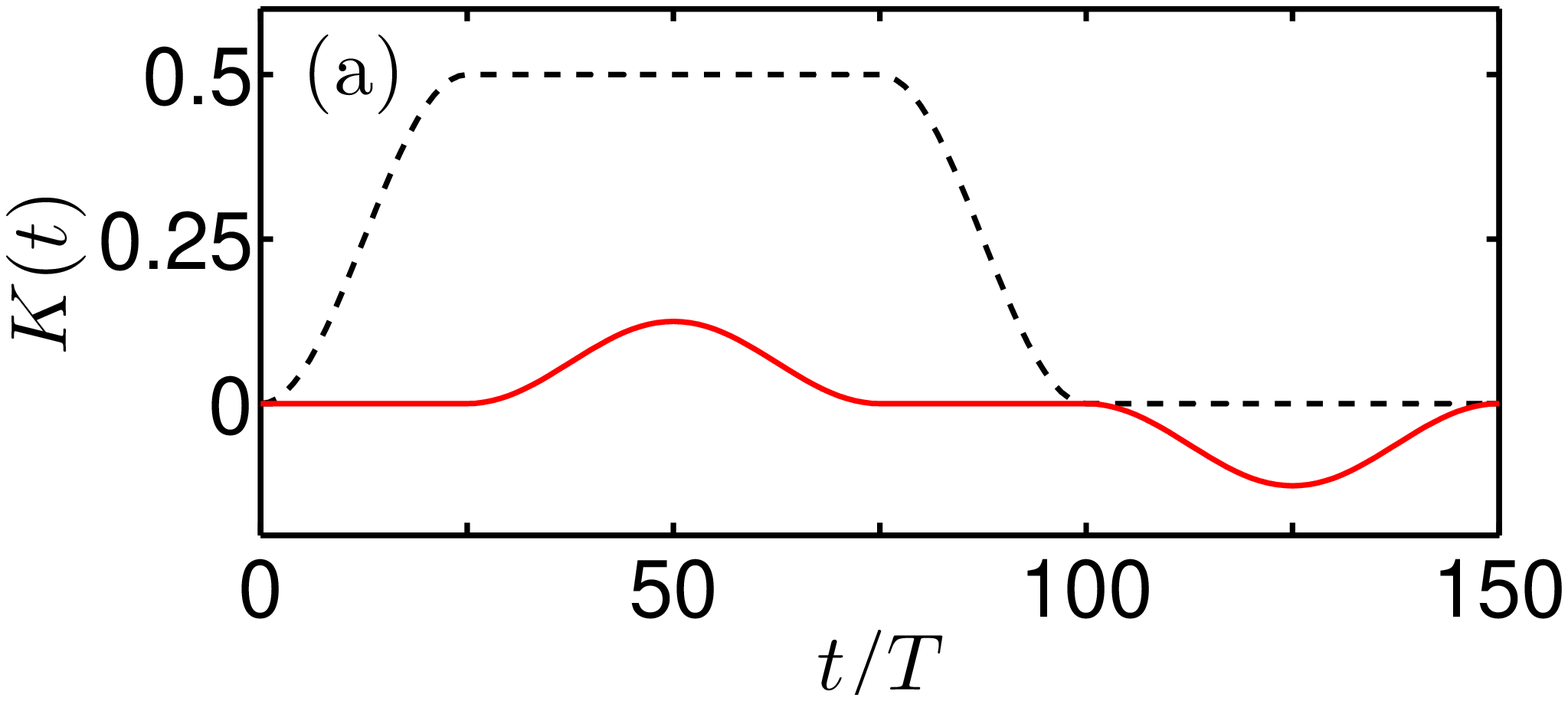}
\includegraphics[width = 0.83\linewidth]{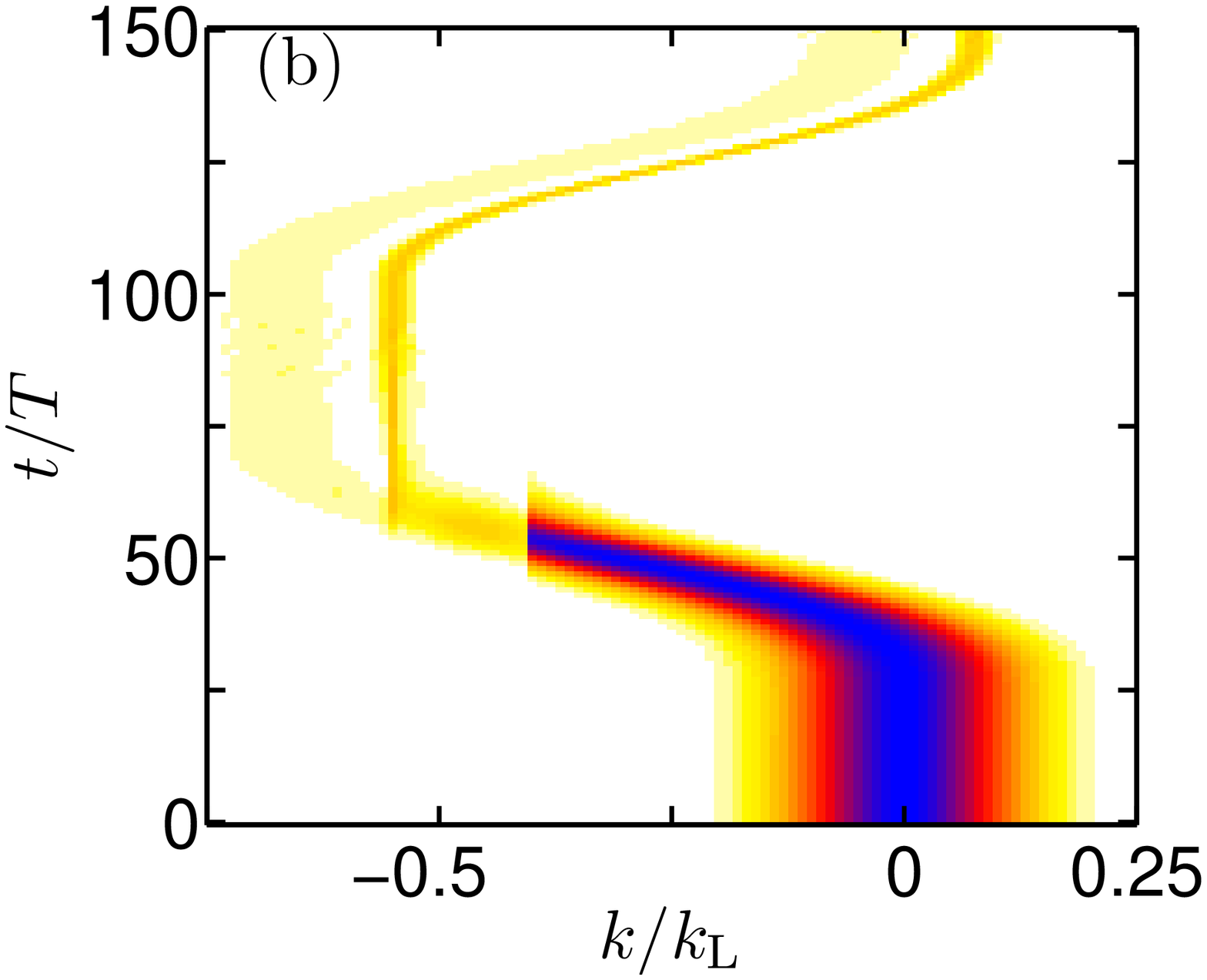}
\includegraphics[width = 0.8\linewidth]{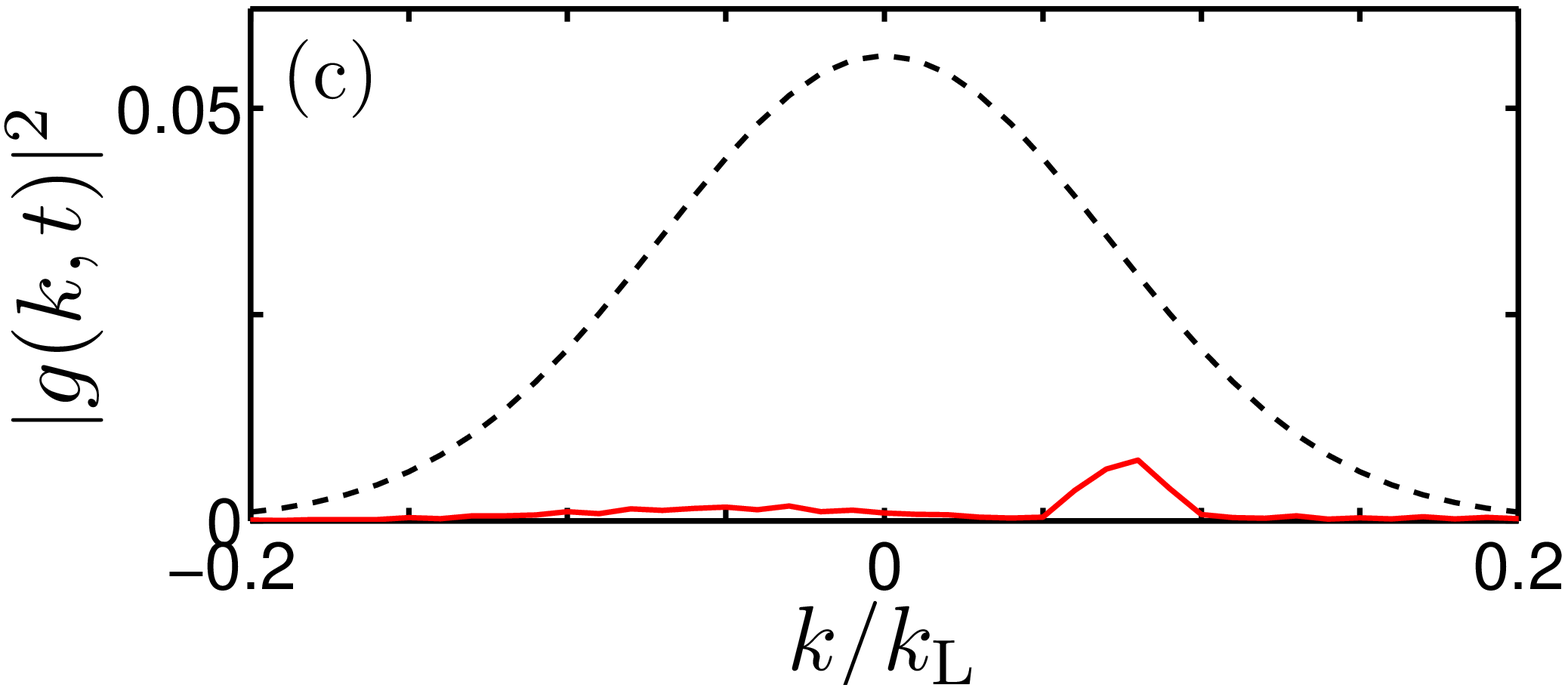}
\caption{(Color online) (a) Protocol for achieving almost complete interband 
	population transfer in a dressed optical lattice by means of a weak 
	probe force. The dashed line is the scaled envelope of the dressing
	ac force. The solid line is the negative, scaled probe force,
	amplified by a factor of~10.
	(b) Resulting wave-packet dynamics in the Floquet representation,
	shown as a contour plot of $|g_1(k,t)|^2$. Under the first action 
	of the probe force the wave packet is shifted in accordance with 
	the generalized acceleration theorem~(\ref{eq:GAT}), until it 
	undergoes Zener-type transitions to other quasienergy bands at 
	the avoided crossings visible in Fig.~\ref{fig:F_1}(b). After 
	the dressing force is switched off, the second, reversed action of 
	the probe force shifts the remaining part of the packet back to the 
	Brillouin zone center.
	(c) Comparison of the initial wave packet (dashed line) with the 
	part of the wave function that remains in the lowest energy band 
	at the end of the process (solid line).}  
\label{fig:F_3}
\end{figure}

At this point an important issue needs to be stressed: The concept of adiabatic
following, or parallel transport in a differential-geometric language, usually 
is applied to energy eigenstates;~\cite{Simon83,Berry84} in the context of
optical lattices this has been exploited, e.g., by Fratalocchi and Assanto
for studying nonlinear adiabatic evolution and emission of coherent Bloch 
waves.~\cite{FratalocchiAssanto07} In contrast, here we consider adiabatic 
following of explicitly time-dependent {\em quasi\/}energy eigenstates, that 
is, of solutions to the quasienergy eigenvalue equation~(\ref{eq:QEE}); this 
is what allows us to separate the fast, oscillating time dependence of the 
driving force from the slow, parametric time dependence of its envelope.    

Having learned these lessons, we now set the generalized acceleration
theorem~(\ref{eq:GAT}) to work. Suppose that we are prompted to empty the 
ground-state energy band. Starting again from an initial wave 
packet~(\ref{eq:INI}), we then may proceed as follows: First we smoothly 
turn on an ac force which dresses the lattice, creating avoided quasienergy 
crossings initially not ``seen'' by the adiabatically following packet. For 
instance, we may wish to utilize the avoided crossings showing up in 
Fig.~\ref{fig:F_1}(b). To this end, we again take an ac force with frequency 
$\omega = 3.71 \, E_{\rm r}/\hbar$ and fix its scaled driving amplitude at 
the plateau value $K = 0.5$. This dressing force is switched on during 25 
cycles with half a squared-sine envelope, maintained at maximum amplitude 
for 50 further cycles, and switched off again for another 25 cycles, as 
sketched in Fig.~\ref{fig:F_3}(a). If this were all we did, 
the wave packet would simply undergo adiabatic evolution and finally restore 
its initial condition, as previously observed in Fig.~\ref{fig:F_2}. Instead, 
once the maximum dressing amplitude has been reached, we now apply an 
additional weak probe force $\Fp (t)$ in order to exploit Eq.~(\ref{eq:GAT}) 
for moving the packet away from the Brillouin zone center, driving it 
over the avoided crossings that have opened up in Fig.~\ref{fig:F_1}(b). 
This probe force is implemented in the form of two smooth, squared-sine 
shaped dc pulses, one acting during the plateau of the dressing pulse, the 
other acting with reversed sign after the dressing pulse is over, as drawn 
in Fig.~\ref{fig:F_3}(a). The maximum strength of the probe force here is 
only 2.5\% of that of the dressing force; for better visibility, the probe 
force is magnified in Fig.~\ref{fig:F_3}(a) by a factor of 10.  

It is now almost obvious how to describe the response of the wave packet
within the Floquet picture: The initial state~(\ref{eq:INI}) first is
adiabatically shifted into a single quasienergy-band packet during the
turn-on of the dressing force. In contrast to a crystal-momentum 
representation, all dressing-induced fast oscillations are taken out 
of the dynamics of $g_1(k,t)$ in the Floquet representation, as shown in 
Fig.~\ref{fig:F_3}(b). When the first probe pulse acts at constant dressing 
amplitude, it forces the wave packet over the avoided crossing seen in 
Fig.~\ref{fig:F_1}(b), so that the packet undergoes Zener-type transitions to 
``higher'' quasienergy bands,~\cite{Zener34,ArlinghausHolthaus10} splitting 
into individual subpackets associated with the different quasienergy bands 
involved. When the dressing force is switched off, each of these subpackets 
moves adiabatically on its own quasienergy surface, finally reaching the 
continuously connected Bloch bands. The second, reversed probe pulse, 
applied after the dressing pulse is over, then acts in accordance with 
Bloch's original acceleration theorem~(\ref{eq:OAT}), shifting the various 
subpackets back to the Brillouin zone center. In the scenario displayed in 
Fig.~\ref{fig:F_3}, the lowest band is almost entirely depopulated by the 
probe-induced Zener transitions, so that only a marginal fraction of the 
initial packet returns, as depicted in Fig.~\ref{fig:F_3}(c). Thus, the main 
part of the initial packet has been placed in higher Bloch bands, as intended.
We have also checked by explicit calculation that without the comparatively
weak probe pulses the returning wave packet would be almost identical to
the initial one.

\begin{figure}[t]
\centering
\includegraphics[width = 0.8\linewidth]{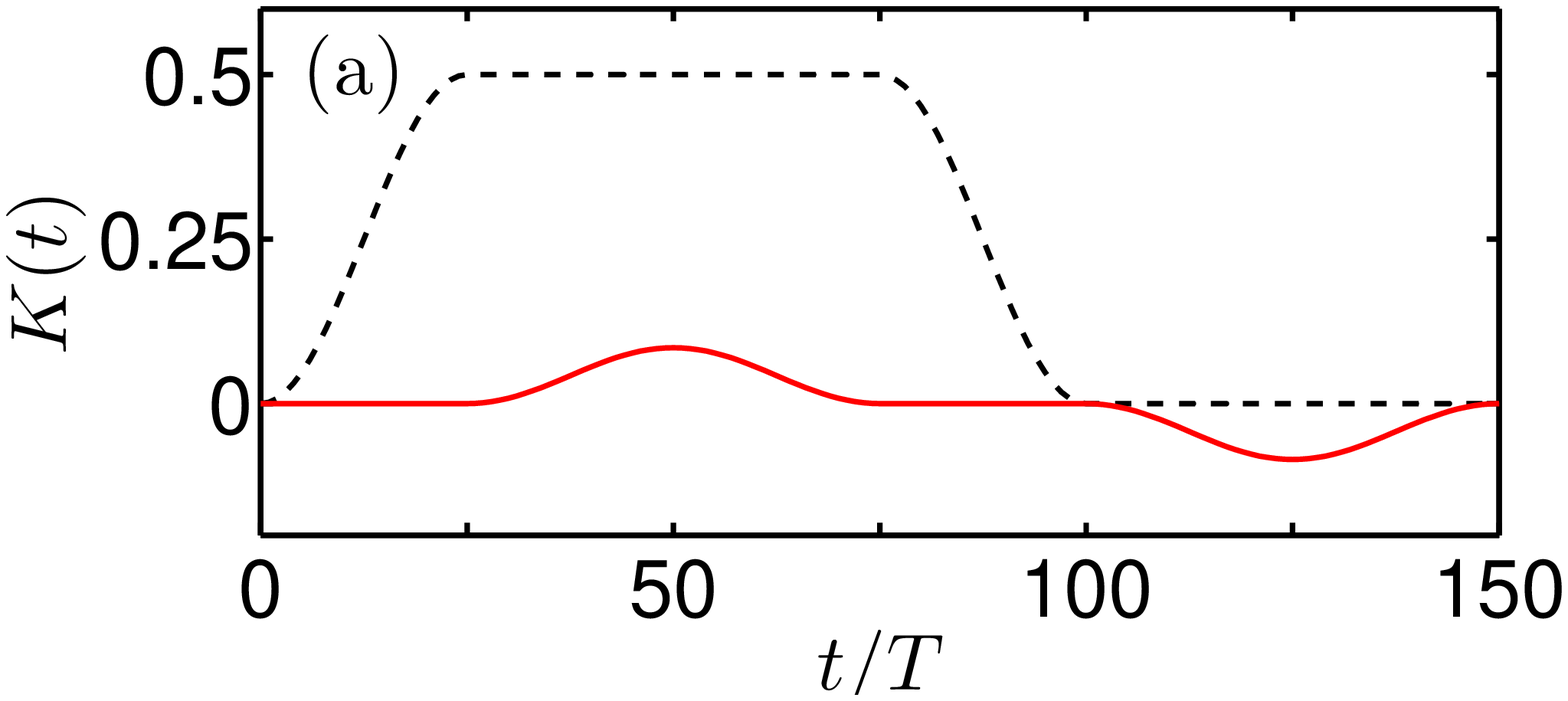}
\includegraphics[width = 0.83\linewidth]{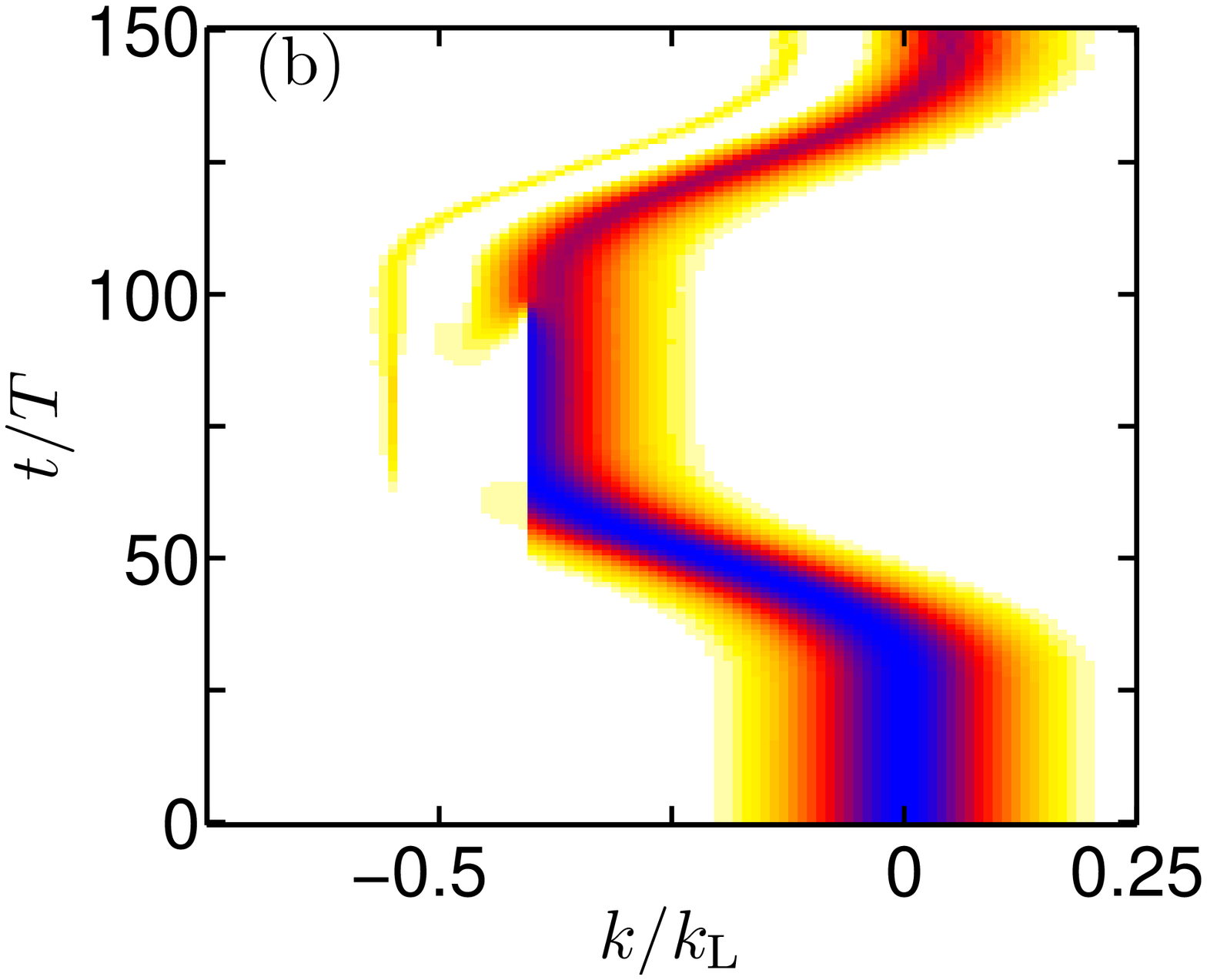}
\includegraphics[width = 0.8\linewidth]{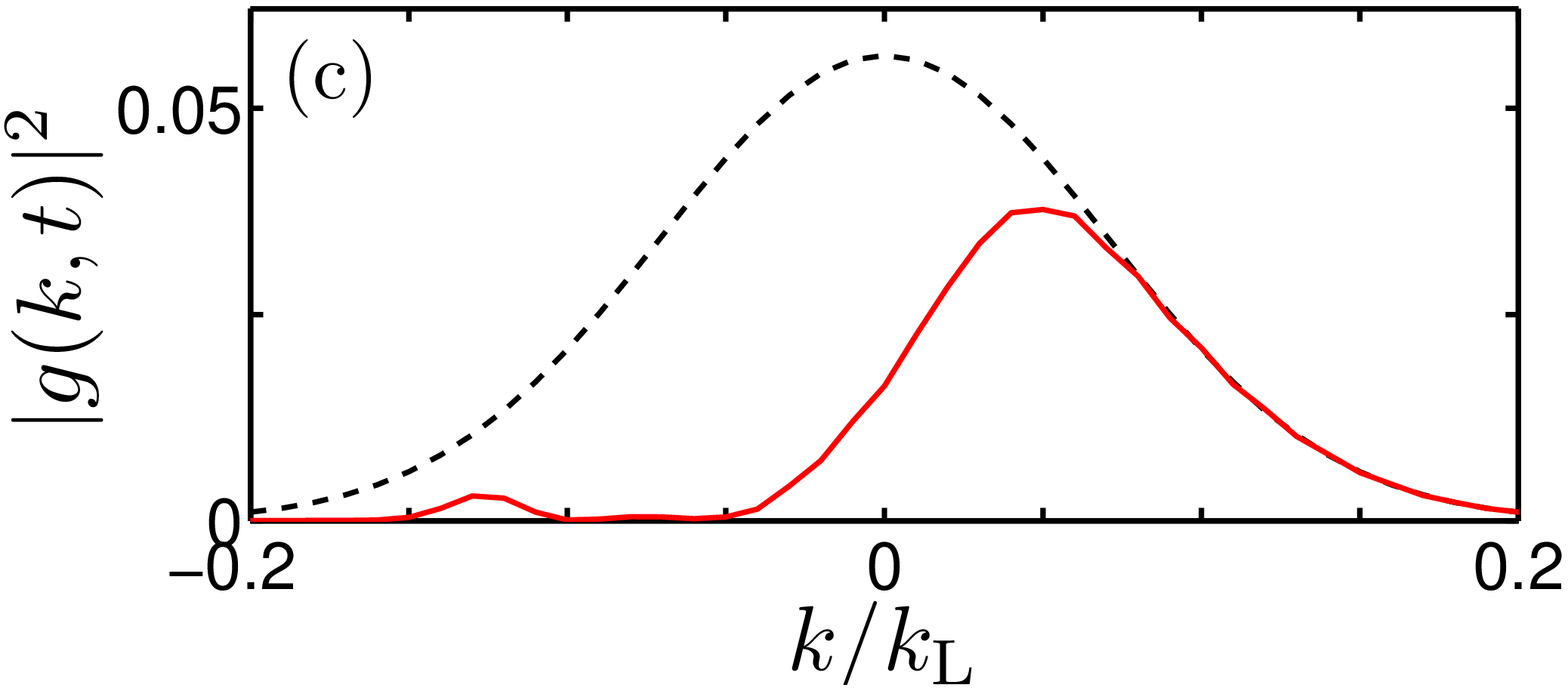}
\caption{(Color online) (a) Protocol for achieving partial interband 
	population transfer in a dressed optical lattice by means of a weak 
	probe force. The dashed line is the scaled envelope of the dressing
	ac force, which is the same as in Fig.~\ref{fig:F_3}. 
	The solid line is the negative, scaled probe force, amplified by a 
	factor of~10; this force is weaker than the one in Fig.~\ref{fig:F_3}.
	(b) Resulting wave-packet dynamics in the Floquet representation,
	shown as a contour plot of $|g_1(k,t)|^2$. Under the first action 
	of the probe force the wave packet is shifted in accordance with 
	the generalized acceleration theorem~(\ref{eq:GAT}), but not as
	far as in Fig.~\ref{fig:F_3}, such that it undergoes only partial 
	Zener transitions. After the dressing force is switched off, 
	the second, reversed action of the probe force shifts the remaining 
	part of the packet back to the Brillouin zone center.
	(c) Comparison of the initial wave packet (dashed line) with 
	the part of the wave function that remains in the lowest energy
	band at the end of the process (solid line).}  
\label{fig:F_4}
\end{figure}

The above example of our ``dressing and probing'' strategy immediately
lends itself to a host of further modifications and extensions. To give
but one further instance, if the probe pulse is still weaker, such that the 
wave packet does not pass over the avoided-crossing regime, but rather stops 
there, the Zener transitions are incomplete, so that a signifcant part of 
the initial state is recovered when the process is over. This is elaborated  
in Fig.~\ref{fig:F_4} with the same dressing force as above, but now the 
maximum strength of the probe force amounts to only 1.7\% of that of the
dressing force. The final subpacket still occupying the lowest Bloch band 
then is no longer centered around $k/k_{\rm L} = 0$, implying that this 
subpacket will move over the lattice. In a sense, the left wing of the initial 
wave packet has been cut out, so that Fig.~\ref{fig:F_4} may be regarded as 
a particular paradigm of ``wave-packet surgery.''~\cite{ArlinghausHolthaus11}

\section{Conclusions}

Summarizing our line of reasoning, we have introduced in Sec.~\ref{sec:S_2}
a representation of wave packets of quantum particles in spatially periodic 
lattices subjected to homogeneous, time-periodic forcing which is based on 
an expansion with respect to spatiotemporal Bloch waves and reduces to 
the standard crystal-momentum representation when the forcing is turned off. 
It embodies forcing-induced oscillations into the basis, so that only the 
actually relevant dynamics remain to be dealt with. Within this Floquet 
representation one encounters many features already familiar from solid-state 
physics in time-independent lattice potentials, but here their scope is 
different. As a prominent example, the generalized acceleration theorem derived
in Sec.~\ref{sec:S_3} takes the same form as its historic antecessor formulated 
by Bloch,~\cite{Bloch28} but applies to single quasienergy band dynamics, 
which can be drastically different from single energy band behavior. There 
are further features which can be carried over from the crystal-momentum 
representation to the Floquet picture and acquire a modified meaning there, 
such as the expression for the group velocity of a wave packet or Zener 
transitions among different bands.

The super Bloch oscillations considered in Sec.~\ref{sec:S_4} provide a mainly
pedagogical example which can be worked out in full detail analytically.
Here the Floquet picture cannot exert its full strength, because one assumes 
\emph{a priori} that the driving force does not induce transitions from the 
initially occupied energy band to other ones, so that the historic acceleration 
theorem remains capable of describing the entire dynamics. The Floquet approach 
leads to exactly the same result, but implies a different viewpoint, separating
the dc component of the force into one part which is resonant with the
ac component, and together with the latter dresses the lattice, creating a 
quasienergy band; the remaining residual part of the dc force then probes 
this new quasienergy band, rather than the original unperturbed energy
band.

This theme of ``dressing and probing'' also prompts far-reaching strategies 
for achieving coherently controlled interband population transfer and even 
more. Two basic examples for this have been discussed in Sec.~\ref{sec:S_5}, 
but the possibilities obviously extend much farther. Utilizing the generalized
acceleration theorem, an initial wave packet may by split coherently into two 
components at an avoided quasienergy band crossing in a dressed lattice, 
and the lattice may then be redressed (that is, exposed to an ac force 
with different parameters) such that another quasienergy band structure is 
generated, possibly involving avoided crossings which affect only one of the 
daughter wave packets created  in the first step, but not the other. Moreover,
daughter wave packets can be made to move, possibly into different directions, 
and to interfere with other wavelets having been manipulated separately before  
in distant parts of the lattice. This vision apparently will be hard to realize
with traditional solids, but it has come into immediate reach in current 
laboratory experiments with weakly interacting Bose-Einstein condensates 
in driven optical lattices. Seen against this background, the generalized 
acceleration theorem almost provides a blueprint for a wave-packet processor.

\begin{acknowledgments}
M.H.\ thanks T.~Monteiro for a thorough discussion of super Bloch oscillations.
This work was supported by the Deutsche Forschungsgemeinschaft under Grant 
No.~HO~1771/6. 
\end{acknowledgments}

\end{document}